\documentstyle[aps,psfig]{revtex}

\begin{document}

\twocolumn[

\begin{center}

{\large\bf Mutual synchronization and clustering in randomly coupled \\
chaotic dynamical networks}\\

\vspace{.5cm}
{Susanna C. Manrubia and Alexander S. Mikhailov}\\
{\small \it Fritz-Haber-Institut der Max-Planck-Gesellschaft,
Faradayweg 4-6, 14195 Berlin, Germany}\\
\end{center}

\vspace{.3cm}
{\small
We introduce and study systems of randomly coupled maps (RCM) where the
relevant parameter is the degree of connectivity in the system. Global
(almost-) synchronized states are found (equivalent to the synchronization
observed in globally coupled maps) until a certain critical threshold for
the connectivity is reached. We further show that not only the average
connectivity, but also the architecture of the couplings is responsible for
the cluster structure observed. We analyse the different phases of the
system and use various correlation measures in order to detect ordered
non-synchronized states. Finally, it is shown that the system displays a
dynamical hierarchical clustering which allows the definition of emerging
graphs.} \\

{PACS number(s): 05.45.+b, 05.20.-y, 05.90.+m}
\vspace{.5cm}
]

\section{Introduction}

Since their introduction in 1989 \cite{Kan,Kan3}, globally coupled maps
(GCM) have turned out to be a paradigmatic example in the study of the
emergent behavior of complex systems as diverse as ecological networks, the
immune system, or neural and cellular networks. It has been shown that
mutual synchronization of chaotic oscillations is a robust property
displayed by GCM \cite{Kan3,Kan2}. Effects of mutual synchronization are
also known for coupled chaotic oscillators with continuous time \cite{Hea}.
A recent study \cite{D1} of large globally coupled populations formed by
chaotic R\"{o}ssler oscillators has revealed that mutual synchronization and
dynamical clustering in these systems are similar to the respective
behaviour found in GCM. Another well investigated class of
self-synchronizing systems represents lattices of locally (e.g.,
diffusively) coupled oscillators \cite{Kur}. Moreover, oscillator systems
with both local and global coupling between elements have also been
discussed \cite{Mikh}. A common property of GCM and other above-mentioned
systems is their architectural symmetry: The pattern of connections of any
individual element is identical. This internal symmetry is preserved in the
fully synchronized dynamical states and spontaneously broken when dynamical
clustering takes place.

The architecture of dynamical networks found in real ecological or other
systems would rarely be so symmetric. Even in the situations with high
connectivity, when the links can extend to many distant elements, the graph
of connections may have a complex topology. The question is therefore
whether and in what form mutual synchronization and dynamical clustering can
persist in such complexely connected networks, lacking a structural
symmetry. In this direction, among the few cases already explored we can
mention a layered system of identical random neural networks with partial
(though regular) connectivity among layers \cite{M1}, coupled lattice maps
with connections extending further than to nearest neighbors \cite{NN}, as
well as coupled lattice oscillators in 2-D with different coupling schemes
\cite{Niebur} or a model ecosystem with partial connectivity \cite{Guido}.
More recently, ensembles of nonlinear oscillators \cite{Stiller} and GCM
\cite{Dam} with random interactions and variable symmetry (still globally
coupled) have also been analyzed.

The analysis of systems formed by few coupled
caotic elements complements the previous approach and has provided a 
better understanding of the collective 
behaviour displayed by large ensembles of regularly connected chaotic 
oscillators. The stability of the synchronous state can be  
in this case easily quantified by means of the transverse Lyapunov
exponent $\lambda_{\perp}$, which changes sign in a blow-out bifurcacion
\cite{SO} and makes the synchronous state unstable. Near the bifurcation, 
on-off intermittency \cite{PST} and riddled basins of attraction \cite{RB} 
are observed. Much attention has been devoted to the dynamical 
properties of systems formed by two coupled, identical chaotic elements 
\cite{FY}. The synchronization properties of two coupled logistic 
maps have been extensively investigated \cite{MMPM}.

As the next step towards understanding the synchronization phenomena in
complex networks, one can consider systems formed by a large number of
identical dynamical
elements that are connected by identical symmetrical links but where the
pattern of connections between elements is random. This is the starting
point of our paper. In the following sections we introduce and analyse what
we have termed Randomly Coupled Maps (RCM), that is, networks of chaotic
maps connected (symmetrically) at random where the relevant parameter is the
average connectivity in the system.

Our main result is that mutual synchronization and dynamical clustering are
possible in RCM even when a significant fraction (up to 40-45\%) of all
potential connections is absent. However, the synchronization and clustering
phenomena in these systems are different in certain aspects from what is
known for GCM. Exact synchronization and the formation of identical
dynamical states of the elements are not found here. Instead, either one 
or several compact clouds (fuzzy clusters) are formed. These fuzzy clusters
are dynamical objects which split into subclusters or join other groups
of elements. Such dynamical hierarchy of clusters is almost never 
completely fixed in time. A closely related effect is that the asymptotic
dynamical behavior in RCM is never sensitive to the initial conditions.
However, the architecture of a particular network may bias the
synchronization process and make certain cluster distributions more
favourable. The role of the network is particularly evident when the
system is small, ($N \simeq O(1)$): In this case, the synchronization
properties of the system are strongly dependent on the particular
architecture, and graphs with the same connectivity might have very
different collective behavior. We will show that only in the thermodynamic 
limit do the synchronization properties of RCM become equivalent 
to the globally coupled case.

In the next Section we introduce the model and describe its dynamical
behavior. In Section III the synchronous and the partially ordered phases of
the system are quantitatively characterized by computing 
properties such as the mutual information and two order parameters of the
synchronization transition. A more detailed statistical investigation is
then performed in Section IV where distributions over pair distances are
constructed. Section V is devoted to the analysis of the emergent cluster
structure in partially condensed phases. A complement to the latter section
is the Appendix, where an example of a small system with varying network
architectures is considered. Finally, in Section VI we discuss our results,
come to the conclusions and outline future extensions of this study.

\section{The Randomly Coupled Maps}

We begin by exactly defining what we have termed Randomly Coupled Maps.
Instead of a globally coupled system, this will be a network of connections
characterized through a random matrix $T_{ij}$, the elements of which are
either $0$ (when a connection between maps $i$ and $j$ is absent) or $1$. In
our analysis, we assume that the matrix is symmetric, i.e. $T_{ij}=T_{ji}$,
and all diagonal elements are set to $0$ ($T_{ii}=0$). An important property
of such random networks is their average connectivity

\begin{equation}
\nu =\frac{1}{N(N-1)}\sum_{i,j=1}^{N}T_{ij}
\end{equation}
Hence, each element will be on the average connected to $\nu (N-1)$ elements
in the system. If $\nu =1$, the system is globally coupled and our system
reduces to this known system.

The collective dynamics of RCM is defined as

\begin{eqnarray}\label{RCM}
x^{i}(t+1)=\left( 1-{\frac{\epsilon }{\nu }}{\frac{1}{{N-1}}}
\sum_{j=1}^{N}T_{ij}\right) f\left( x^{i}(t)\right) \nonumber \\
+ {\frac{\epsilon }{\nu }}
{\frac{1}{{N-1}}}\sum_{j=1}^{N}T_{ij}f\left( x^{j}(t)\right)
\end{eqnarray}
where $\epsilon $ specifies the strength of the coupling and $f(x)$ is the
individual map. This collective dynamics can thus be understood as involving
diffusion on a graph: Each node in the random graph diffuses a certain
fraction of material (its state) to the elements to which it is connected,
and receives a comparable influence from them. The intensity of coupling
between the network elements is specified by the parameter $\epsilon $ (we have
the constraint $\epsilon <\nu $). Nevertheless, the combination $\epsilon
/\nu$, which can be interpreted as a diffusion coefficient, plays an important 
role. Note that in the limit $\nu N \to \infty$ all RCM with a given
connectivity $\nu$ become statistically identical. Indeed, in this limit
each element will have the same number $\nu N$ of connections, although
still it is linked only to a randomly chosen subpopulation. If the
mean-field approximation holds, the behavior of this system would
therefore be equivalent to the behavior of a GCM with coupling intensity 
$\epsilon$.

In this paper we investigate differences in the behavior of GCM and RCM
with large, but finite numbers of elements. We work with the logistic map 
$f(x)=1-ax^{2}$ and use
values of $a$ such that the dynamics of a single map is chaotic. 
Our numerical simulations are performed for graphs with sizes up to 
$N=2048$
elements and with connectivities in the interval $0.5<\nu \le 1$. Such
graphs have been randomly generated by independently choosing every possible
connection with the same fixed probability. We have checked that thus
generated graphs remained fully connected, i.e. they could not be further
separated into two disconnected parts. 

Considering that we wish to compare the behaviour of RCM with that of GCM,
it is of interest to begin by briefly recalling the dynamic behaviour of GCM
when the coupling strength $\epsilon $ is decreased from 1 to 0 (cf. \cite
{Kan3}). When, for example, $a=2$, the synchronous
phase in GCM is maintained until $\epsilon =0.5$, where it destabilizes and
is substituted by the so called `glassy phase'. In Kaneko's terminology,
`glassy' means that the final attractor is sensitive to the initial state of
the maps, and thus a multiplicity of attractors is to be found in this phase
for the same parameters value. For $\epsilon \approx 0.33$, an ordered phase
sets in: The sensitivity to the initial conditions disappears and the
elements group together in only 2 or 3 different clusters. A narrow
intermittent band in $0.21<\epsilon <0.25$ preceeds the turbulent phase, in
which the number of clusters in all the attractors is of order $O(N)$.

We first examine the possibility of full mutual synchronization in RCM. Such
full synchronization takes place in GCM when the coupling strength exceeds a
critical value \cite{Kan2}. In the fully synchronized (coherent) regime the
states of all maps in GCM are identical. Our analysis reveals that such
exact synchronization does not occur in RCM. However, at sufficiently high
coupling intensities, all elements move together in a single compact cloud
(a fuzzy cluster), so that the typical distances between their trajectories
remain below $|x^{i}(t)-x^{j}(t)|\le 10^{-8}$ (this is the single precision
for real numbers in our computer). The fuzzy synchronous phase is maintained
for a certain range of values of the coupling strength $\epsilon $, and then
a sudden transition to an asynchronous state (see the discussion below) is
observed. Fig. 1 shows how this transition proceeds for a randomly chosen
network with $\nu =0.8$ and $N=50$ elements. For comparison, we show in the
same figure the respective behaviour of GCM. While for GCM the
turbulent-ordered-glassy-synchronous sequence of phases is clearly seen
when $\epsilon $ decreases, only the synchronization breakdown for RCM is
apparent.

\begin{figure}
\centerline{\psfig{file=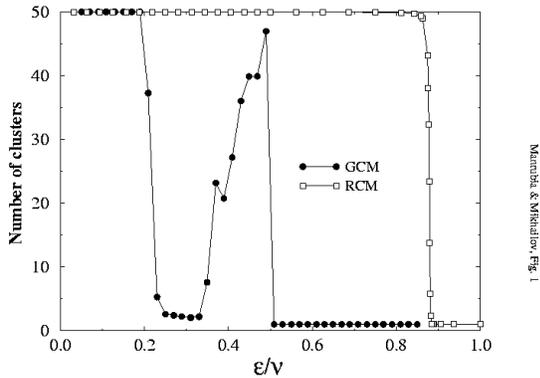,width=8.cm,angle=0}}
\caption{Numbers of clusters as functions of the coupling strength $\epsilon$
for RCM of size $N=50$ and connectivity $\nu =0.8$ and for GCM ($\nu=1$) of 
size $N=50$. In the RCM case, two elements $i$ and $j$ are considered to 
belong to the same cluster if $|x^{i}(t)-x^{j}(t)|<10^{-8}$. In the 
GCM case, two elements belong to the same cluster if their states are 
identical.}
\end{figure}

We have studied how the synchronization threshold in RCM depends on the mean
connectivity $\nu$ of the network and on its size $N$. We accept that the 
elements in a given network are synchronized if the average distance $\bar d$
between pairs of elements during an interval $\Delta t$ is smaller than
$10^{-8}$, after a transient is discarded, that is

\begin{equation}\label{cond}
{\bar d}= {1 \over N (N-1) \Delta t}
\sum_{\langle ij \rangle} \sum_{\Delta t} |x^i(t) - x^j(t)| \;\;\; < \;
10^{-8}.
\end{equation}
The simulations begin at a value of $\epsilon<\epsilon^*$ which is 
stepwise increased in amounts $10^{-3}$ until the condition
(\ref{cond}) is fulfilled.
We have also observed that immediately prior to the synchronization
transition, the distance $\bar d$ undergoes a sharp cutoff and
afterwards stabilizes around small values which depend on the system 
size ($\bar d$ is in the synchronous phase a decreasing function of $N$).
In Fig. 2 the distance to the synchronization threshold of GCM is shown
as a function of the network size when the connectivity is kept constant
($\nu =0.9$). Numerically, we obtain 

\begin{equation}
\epsilon^* - \epsilon_{GCM}^*  \simeq {1 \over \sqrt
N} \; \; .
\end{equation}
Large open circles are averages over 10 to 30 independent graphs, and are 
represented together with the dispersion

$$\Delta \epsilon^* (N) = \left[
{\sum_j (\epsilon^*_j(N)-\bar{\epsilon}^*(N))^2 \over (N(N-1))} \right]^{1/2}$$ 
in the thresholds (error bars). The index $j$ stands for different 
networks formed by the same number of elements $N$, while $\bar{\epsilon^*}(N)$
is the average value for each size. Small solid circles represent the
synchronization threshold for each of these graphs. The error in the
determination of $\epsilon^*$ for each network is $10^{-3}$. The solid
line in the main plot has slope $-1/2$.

\begin{figure}
\centerline{\psfig{file=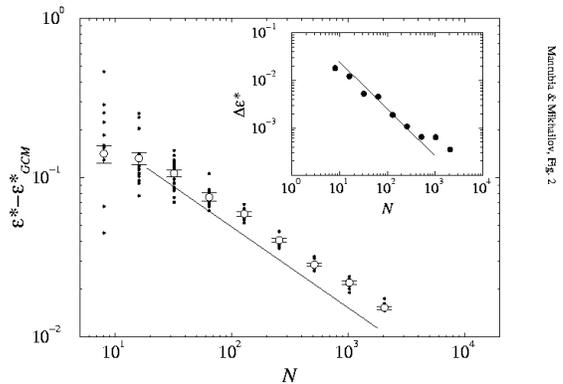,width=8.cm,angle=0}}
\caption{Dependence of the synchronization threshold $\epsilon ^{*}$
on the network size $N$. The open circles correspond to average values of 
the critical coupling intensity $\epsilon^{*}$ for 
networks of sizes ranging from $N=2^3$ to $2^{11}$ with the same 
connectivity $\nu =0.9$. The solid line has slope $-1/2$. In the insert
we show the dispersion in the values of $\epsilon ^{*}$ as a function of
the system size. The solid line has slope $-1$. In this case, $a=2$ and thus
$\epsilon^*_{GCM}=0.5$.}
\end{figure}

As larger networks are considered, the dispersion $\Delta \epsilon^* (N)$
in the synchronization thresholds for different networks with the same
connectivity becomes smaller (insert). Our numerical results point to a
dependence of the form $\Delta \epsilon^* \simeq N^{-1}$. The fact that
the dispersion in the values of $\epsilon^*$ tends to zero in the limit
$N \to \infty$ for a fixed $\nu$ indicates that RCM are characterized by 
self-averaging quantities.
In view of these results, the RCM should be well
described by a mean-field approximation, thus by GCM, in the limit 
$\nu N \to \infty$. In fact, 
the numerical simulations agree with this picture.
In the opposite limit, when the number of elements in the systems is small
($N \simeq O(1)$), the threshold at which the group first synchronizes is 
very sensitive to the particular way the maps are connected, as can be
already seen for $N \le 64$ in Fig. 2 (see also the Appendix).
Fig 3. depicts the synchronization threshold for systems with $N=50$ and 
$N=500$. Every point corresponds to a fixed network with connectivity 
given in the $x-$axes and the critical value $\epsilon^*$ in the $y-$axes. 

\begin{figure}
\centerline{\psfig{file=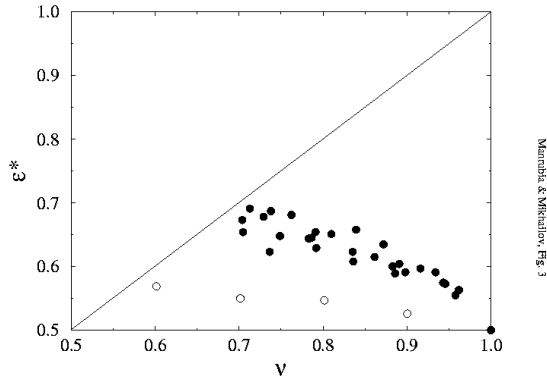,width=8.cm,angle=0}}
\caption{Threshold to synchronization. Small solid circles are critical
thresholds for systems of size $N=50$ and big open ones correspond to 
$N=500$. We restrict ourselves to values $\epsilon \le \nu$, as
discussed in the main text. The line $\epsilon=1/2$ is the stability 
threshold for GCM, and acts as a lower boundary for RCM.}
\end{figure}

\begin{figure}
\centerline{\psfig{file=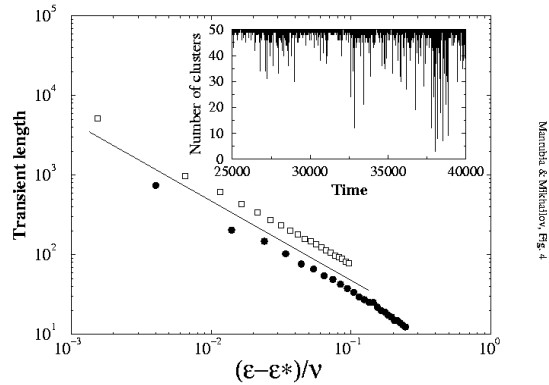,width=8.cm,angle=0}}
\caption{Divergence in the transient time $t_{c}$ when approaching the
critical point $\epsilon ^{*}$ for a randomly chosen network with $N=50$
and $\nu =0.8$ (squares) and for the GCM of the same size (filled circles).
The solid line indicates the divergence law with $\tau =1$. The inset
displays the number of clusters as function of time for the RCM at $\epsilon
=0.7$.}
\end{figure}

The synchronization threshold for GCM has previously been exactly 
determined \cite{Kan3}. 
When $N \to \infty$ it is given by the condition $\epsilon=\epsilon^*_{GCM}$
where the critical coupling is obtained from $\lambda + \log (1- 
\epsilon^*_{GCM})<0$ and $\lambda$ is the Lyapunov exponent of the single 
logistic map. It can be expected that the 
synchronization threshold for RCM approaches the limit 
$\epsilon=\epsilon^*_{GCM}$ when $\nu \to 1$, for any value of $N$, and 
also when $\nu N \to \infty$, and independently of $\nu$, as has been 
discussed. It can be seen from the numerical results represented in Figs. 2 
and 3 that the value $\epsilon=0.5$ (for $a=2$) yields indeed a 
lower estimate for the synchronization threshold in RCM.

The system falls into the synchronous phase after a transient of diverging
length when approaching the synchronization threshold $\epsilon ^{*}$. 
Fig. 4 shows the
dependence of the transient time $t_{c}$ on the distance $\epsilon -\epsilon
^{*} $ to that threshold. The insert shows the dynamics
during a typical transient. We see that strong fluctuations (intermittency)
are accompanying the convergence process. Representing the dependence of the
transient time in the form $t_{c}\propto (\epsilon -\epsilon ^{*})^{-\tau },$
we find that the exponent $\tau \approx 1$ is typical both for RCM and GCM.
More precisely, a least squares fit to numerical data returns 
$\tau_{GCM}=0.997(3)$ and $\tau _{RCM}=0.982(3)$, and the critical threshold 
values to synchronization are $\epsilon _{GCM}^{*}=0.5$ and 
$\epsilon_{RCM}^{*}/\nu \approx 0.87$ for the particular graph of Fig. 4.

Below the synchronization transition, the glassy phase is observed in GCM.
The dynamical behavior of RCM in the region $\epsilon <\epsilon ^{*}$ is
investigated in the next section.

\section{Mutual information and the two order parameters}

We have used three different measures of correlations among elements to
check if the phase with $\epsilon <\epsilon ^{*}$ has still some intrinsic
order . The first of them is the {\it mutual information} between two maps $%
i $ and $j$, $I^{ij}(\Sigma )$. To define this quantity, we introduce a
partition of the phase space of the logistic map in the following way. If
the state of the chosen element $i$ is $x^{i}(t)\ge 0$ then it will be
assigned a value 1, and 0 if $x^{i}(t)<0$. This generates a sequence of bits
in a certain time interval $S_{t}^{i}\in \Sigma \equiv \{0,1\}$ which allows
the calculation of the Boltzmann entropy for the $i$th map \cite{part},

\begin{equation}  \label{bolz}
H^i(\Sigma)=-\sum_{S^i_t=0,1} P(S^i_t) \ln P(S^i_t).
\end{equation}
In a similar way we define the joint entropy for each pair of maps,

\begin{equation}  \label{bolz2}
H^{ij}(\Sigma)=-\sum_{S^i_t=0,1}\sum_{S^j_{t^{\prime}}=0,1} P(S^i_t,
S^j_{t^{\prime}}) \ln P(S^i_t,S^j_{t^{\prime}})
\end{equation}
and finally the mutual information for $i$ and $j$ is given by

\begin{equation}
I^{ij}(\Sigma )=H^{i}(\Sigma )+H^{j}(\Sigma )-H^{ij}(\Sigma ).  \label{inf}
\end{equation}

The mutual information is a good measure of correlations, e.g. it achieves
maximal values near critical points \cite{Com}. In a context closer to ours
it has been shown to accurately discriminate among the different phases of
GCM \cite{JD}. An interesting property of $I^{ij}$ is that it is practically
precision-independent, due to the rough coarse-graining of the dynamics.

As an illustration of the sensitivity of this measure, let us discuss 
which values of 
$I^{ij}$ are expected in two dynamically opposite regimes, i.e for the
synchronous and the turbulent phases. In the coherent (synchronous) phase
where the states of all elements are identical, the sequences $S_{t}^{i}$
are the same for all of them, therefore $H^{i}(\Sigma )=H^{j}(\Sigma )$.
In addition, since the two chosen maps are visiting the same points, we 
have $H^{ij}=H^{i}$ and thus $I^{ij}=H^{i}$, reflecting the trivial 
nature of the correlations. When $a=2$, because of the symmetry in the 
invariant measure of the logistic
map for this parameter (and noting that the synchronized system is equivalent 
to the single map) the mutual information achieves its maximum value.
In fact, under these conditions, $P(0)=P(1)=1/2$, thus maximizing
$H^{i}=\ln 2$ and also $I^{ij}=\ln 2$. For
parameter values $a \ne 2$ the invariant measure is not symmetric around
$0$, $P(0) \ne P(1)$ and hence $I^{ij} < \ln 2$ typically. Nonetheless, as can
be seen in Fig. 5, the synchronous state is clearly detected through this
measure. In the
turbulent phase (assuming that the elements behave independently and are not
correlated) we will have again $H^{i}(\Sigma )=H^{j}(\Sigma )$, but now the
joint probabilities factorize, $P(S_{t}^{i},S_{t^{\prime
}}^{j})=P(S_{t}^{i})P(S_{t^{\prime }}^{j})$ and $H^{ij}=2H^{i}$. In this
phase we therefore expect $I^{ij}\approx 0$ irrespectively of the parameter 
$a$. In the intermediate cases,
where some correlations are present, the mutual information should
take values between the former two limits, $0 \le I^{ij} \le \ln 2$.

To determine the mean mutual information 

\begin{equation}
\langle I^{ij}\rangle = {2 \over N(N-1)} \sum_{\langle i<j \rangle} 
I^{ij}
\end{equation} 
for a network with a given matrix $T_{ij},$ we take an average of the
mutual informations for all possible pairs $(i,j)$ for long enough 
sequences to ensure the stability of the probabilities (typically $\Delta
t=10^{3}-10^{4}$ after discarding a transient). Fig. 5 shows the typical
computed dependences of the mean mutual information $\langle I^{ij}\rangle $
on the coupling strength $\epsilon$ for RCM in the cases $a=2, \; 1.8,$
and $1.6$, and GCM. At low coupling
intensities, the mutual information is zero, indicating the absence of
correlations in the turbulent phase. It starts then to increase and reaches
a maximum. In GCM this maximum corresponds to the ordered phase. When the
coupling intensity is further increased, the mutual information falls down
--at the onset of the glassy phase for GCM-- before it finally increases 
and reaches a stable high value in the synchronous state. 
The effect of a decreased connectivity
translates into a shift of the phases to higher values of the coupling
$\epsilon$. For low enough $\nu$, synchronization is no longer possible
(see also Fig. 3).

Thus, though by direct counting of the number of clusters (Fig. 1) we could
not see any ordering in RCM for $\epsilon <\epsilon ^{*},$ the present
analysis based on the measure of the mutual information clearly shows that the
networks have intrinsic dynamical organization also in this region.

\begin{figure}
\centerline{\psfig{file=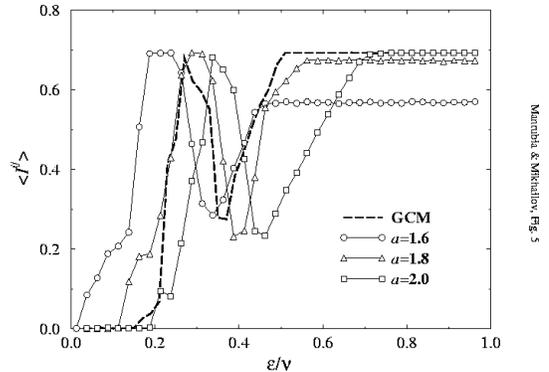,width=8.cm,angle=0}}
\caption{Mean mutual information $\left\langle I^{ij}\right\rangle $ as
function of the coupling intensity $\epsilon$ for a GCM ($\nu=1$) of
size $N=200$ (dashed line) and three randomly chosen networks
of the same size, connectivity $\nu =0.8$ and values of $a$ as shown. 
The sequences $S_{t}^{i}$ were taken after discarding a transient of 
$10^{3}$ steps. The GCM data is additionally averaged over 100 random 
initial conditions, and the RCM over 25 independent graphs.}
\end{figure}

Further characterization of different RCM phases is provided by two {\it
order parameters}\cite{D1}. We examine all different $N(N-1)/2$ pairs of
elements in the system and count how many of them are at a distance shorter
than a fixed given precision $\delta $. The first order parameter 
$r(\delta)$ is defined as

\begin{equation}
r(\delta )={\frac{2}{N(N-1)}}\sum_{\langle i<j\rangle }\Theta \left( \delta
-|x^{i}(t)-x^{j}(t)|\right)  \label{r}
\end{equation}
where $\Theta (x)$ is a step function, $\Theta (x)=0$ for $x<0$ and $\Theta
(x)=1$ for $x>0.$ The sum is taken over all possible ordered pairs $\langle
i<j\rangle $. The second order parameter $s(\delta )$ is given by the
relative number of elements having at least one other element at a distance 
$d<\delta $.

In globally coupled systems the synchronization proceeeds until all the
elements in the same cluster asymptotically reach identical dynamical
states. In this case, one can choose the highest available precision in the
calculation of the two order parameters (in actual simulations it is limited
by the computer precision). As we have already noted, this absolute
synchronization does not occur in RCM. Instead, only clusters of elements 
having close dynamical states are formed here. Therefore the choice of the 
precision $\delta $ becomes important when networks are considered.

We have computed the order parameters as functions of the effective coupling
strength $\epsilon /\nu $ for RCM and the respective GCM using varying
precisions $\delta $. It was found that the results for GCM only weakly
depended on the precision in a wide interval $10^{-10}<\delta <10^{-3}.$ The
curves shown in Fig. 6(a) for GCM have been calculated using $\delta
=10^{-6}. $ We see that the order parameter $r$ for GCM reaches, as should
be expected, the value $r=1$ in the fully synchronous (coherent) state at
high coupling intensities. However, a large relative number $r$ of
synchronous pairs is also found in this case at lower coupling intensities 
$\epsilon $ in the ordered phase. Moreover, the second order parameter $s$ in
this phase is close to 1 indicating that almost all elements belong to one
of the synchronous clusters.

In contrast to globally coupled systems, dynamical clustering and
synchronization in RCM is best resolved when an optimal small precision is
employed. The plots shown for RCM in Fig. 6(b) have been therefore
constructed for the optimal precision $\delta =10^{-3}$. We see that in the
region $\epsilon >\epsilon ^{*},$ where fuzzy mutual synchronization is
observed, both order parameters reach their maximal possible values, $r=s=1.$
Below the synchronization threshold, both order parameters rapidly decrease
but then show a maximum. For small coupling intensities, the order
parameters become very small (a somewhat larger initial level of $s$ is
explained by the fact that $\delta $ is larger here than in Fig. 6(a) and
therefore a small number of pairs separated by the distance $\delta $ is
found already in the turbulent phase, for the random independent
distribution over the one-particle attractor).

\begin{figure}
\centerline{\psfig{file=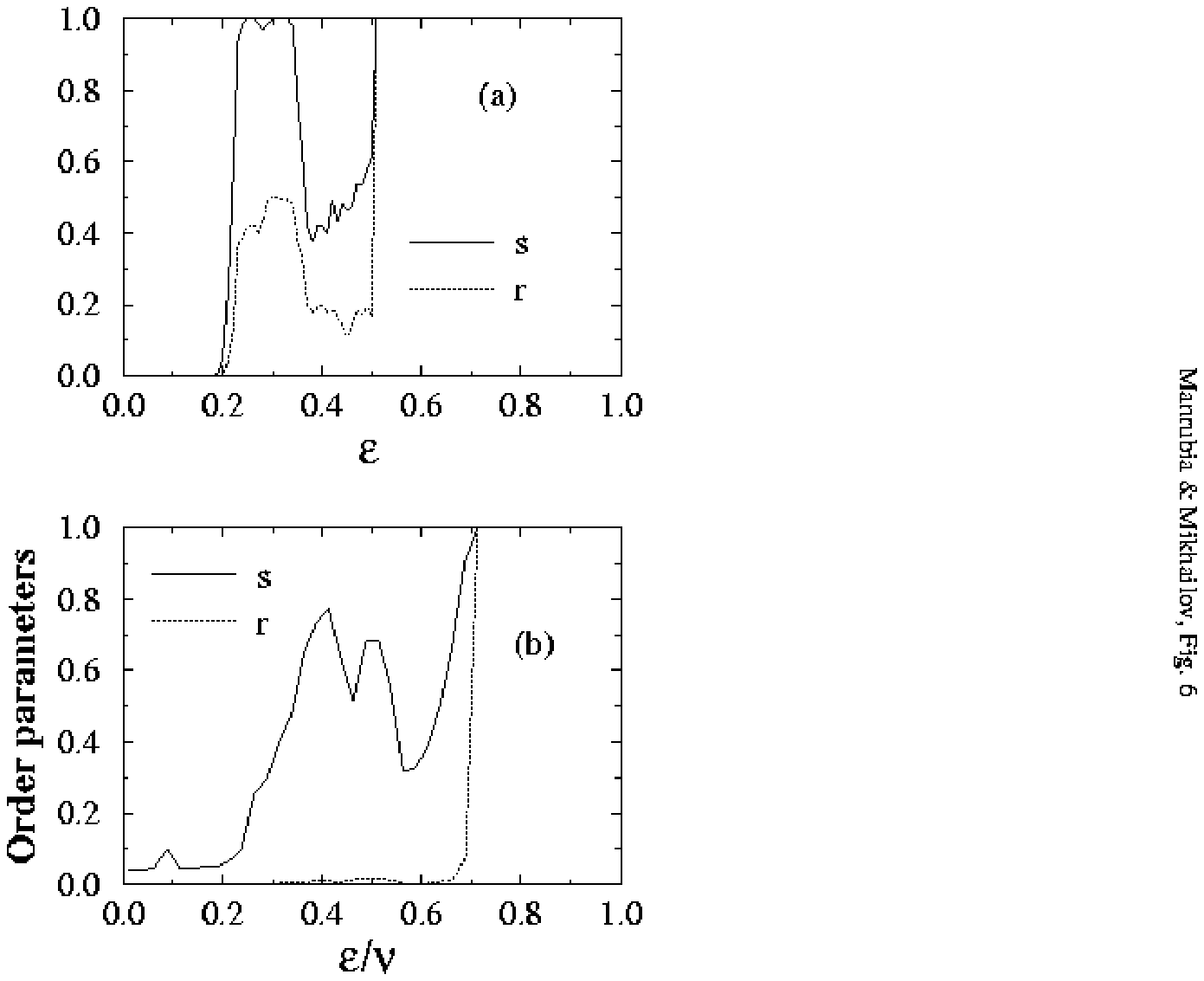,width=8.cm,angle=0}}
\caption{Order parameters $r$ (dotted lines) and $s$ (solid lines) as
functions of the coupling intensity $\epsilon $ for a GCM of size 
$N=250$ (a) and for a randomly chosen network (the same as in Fig. 5)
of size $N=250$ and connectivity $\nu =0.8$ (b). The employed
precisions are $\delta _{GCM}=10^{-6}$ and $\delta _{RCM}=10^{-3}$.
Averaging over 100 random initial conditions is additionally performed for
GCM.}
\end{figure}

Thus, the behavior of the order parameters is again qualitatively similar in
RCM and GCM. We can conclude that within a certain interval of coupling
intensities, dynamical clustering of elements occurs in these systems. The
difference is that, in the case of RCM, the clusters are fuzzy and can
therefore be identified only when a sufficiently low precision is used. One
further difference is that for RCM the order parameters do not fall down so
sharply immediately below the synchronization threshold and a significant
number of elements still has close neighbours in this region.

\section{Distributions over pair distances and dynamical clusters}

Additional information about the structure of different phases in RCM is
provided by histograms of distributions over pair distances. Such histograms
are constructed by counting at a given time moment the numbers of pairs 
$(i,j)$ with distances $d_{ij}=\left| x_{i}-x_{j}\right| $ lying within
subsequent small intervals $\Delta d$. Fig. 7 shows these normalized
histograms for one fixed randomly chosen network with a large number of
elements ($N=1000$) at several coupling intensities $\epsilon $. In the
turbulent phase, a flat distribution corresponding to almost independent
elements is found (Fig. 7(a)). When the coupling intensity is increased, some
inhomogeneities start to develop in the distribution (the mutual information
and the order parameters also begin here to increase). The peaks appearing
later in the distributions indicate the onset of dynamical clustering (Fig.
7(b)). In the situation shown in Fig. 7(c) the system has two fuzzy clusters.
When the coupling intensity is further increased, the cluster structure
is destroyed and the distribution characterized by a broad maximum at
zero distance between elements is formed (Fig. 7(d)). As the coupling
intensity $\epsilon $ grows, this maximum gets increasingly narrow until the
synchronous state is reached at $\epsilon =\epsilon ^{*}$.

\begin{figure}
\centerline{\psfig{file=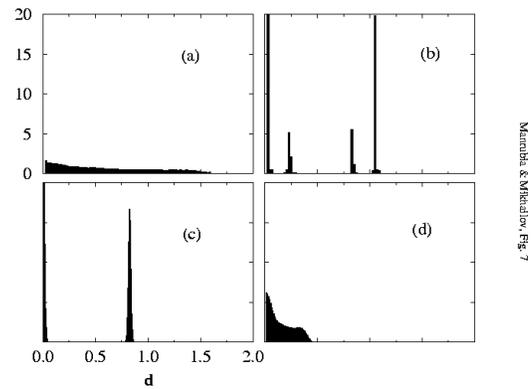,width=8.cm,angle=0}}
\caption{Normalized histograms of distributions over pair distances $d_{ij}$
for different coupling intensities (a) $\epsilon =0.1$, (b) 
$\epsilon =0.25$, (c) $\epsilon =0.35$, (d) $\epsilon =0.45$ for
a randomly chosen network of size $N=1000$ with connectivity $\nu =0.8$.
The histograms are obtained by counting the numbers of pairs with the
distances falling inside subsequent intervals of width $\Delta d=0.01$ at a
fixed time moment $t=200$ after the transient. The vertical and 
horizontal axes have the same scales in all these histograms and stand
for the probability density and for the distance between states, 
respectively.}
\end{figure}

The distributions over pair distances in Fig. 7 correspond to fixed time
moments and therefore cannot tell us anything about dynamical
properties of the clusters. To analyze the underlying dynamical behavior of
the system, we have plotted in Fig. 8(a-d) the typical time evolutions of
the distance between two elements for the histograms displayed in Fig.
7(a-d).

In the turbulent phase (Fig. 7(a) and 8(a)) the pair distance evolves in an
irregular way and shows large variations, as it can be expected for two
independent logistic maps. At the beginning of the clustering phase (Fig.
7(b) and 8(b)), the elements tend to stay much closer (notice the change 
in the vertical scale) and weak aperiodic oscillations are observed. 

For coupling intensities near the top value of the mutual information
(Fig. 7(c) and 8(c)), the elements lock into periodic trajectories. 
Examining the
trajectories of individual elements, we have seen in this case that all of
them are now periodic, though different for different elements of the
system. Thus, the system acquires rigid internal organization and falls into
a state of frozen disorder. The pair distance between two elements in a
cluster shows in this case purely periodic variation. The
clusters are rigid and no exchanges among them are observed. Figures
7(b,c) represent two elements belonging to the same cluster.
The distance between intercluster pairs shows analogous behaviour (i.e.
weak aperiodic oscillations or periodic dynamics) although then the
typical separations are of order $O(1)$. 

An interesting dynamical behavior is observed for higher coupling
intensities, preceeding the synchronization transition (Fig. 7(d)). Now the
elements alternate between short periods of partial synchronization and
excursions away from the incipient cluster (Fig. 8(d), note the increase in
the vertical scale). This form of behavior is in fact very reminiscent of 
on-off intermittency \cite{PST}. 
Such intermittency can explain the origin of the broad shoulder in the
histogram of Fig. 7(d): It is formed by the elements that temporarily find
themselves during a large excursion from the central cluster.

\begin{figure}
\centerline{\psfig{file=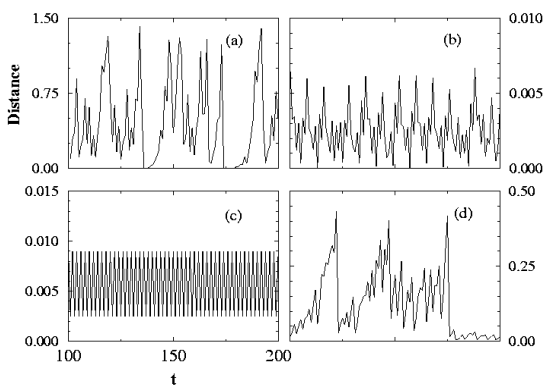,width=8.cm,angle=0}}
\caption{Time evolution of the distance between a pair of elements in the
same random network as in Fig. 7 for different coupling intensities (a) 
$\epsilon =0.1$, (b) $\epsilon =0.25$, (c) $\epsilon =0.35$,
and (d) $\epsilon =0.45.$ 
In cases (b) and (c) the elements belong to the 
same cluster.}
\end{figure}

The previous analysis has been also carried out for $a=1.8$. The above
described picture is also obtained for this other value of the parameter 
in the logistic map and for corresponding coupling strenghts $\epsilon=0.1, 
\; 0.15, \; 0.25$ and $0.29$. This sequence is coherent with the different 
phases detected by the mean mutual information $\langle I^{ij} \rangle$ 
and represented in Fig. 5.

The formation of clusters with periodic dynamics has previously been
observed in the ordered phase of globally coupled logistic maps \cite{Kan2}.
As the coupling intensity is further increased, this ordered phase is
replaced in this system by the glassy phase where the system has a large
number of different attractors and its asymptotic dynamics strongly depend
on the initial conditions. The glassy phase of GCM preceeds the final
transition to the fully synchronous coherent state.

An important result of our study is that the glassy behaviour was absent in
the studied randomly coupled maps. When dynamical clustering was observed in
this system, the cluster structure did not depend on the initial conditions
and was completely determined by the architecture of the underlying graph.
Moreover, the phase of dynamical clustering is separated in RCM from the
synchronous phase by the region of intermittent regimes.

Our interpretation of this finding is that the quenched disorder introduced
by randomly deleting some connections --transforming the GCM into RCM-- 
is to a certain extent equivalent to the introduction of a small amount of 
dynamical noise (either multiplicative or additive) in a globally coupled 
system \cite{com}. 
We have checked this conjecture by constructing the distributions 
over pair distances for GCM and RCM in the glassy and intermittent phases,
respectively. Moreover, we have also computed similar distributions for GCM
where an additive or a multiplicative noise have been included. The dynamical
evolution of the noisy GCM is defined through

\begin{equation}
x^i(t+1)= (1-\epsilon) f(x^i(t)) + {\frac{\epsilon }{N}} \sum_{j=1}^N
f(x^j(t)) + \zeta g(x^i(t)),
\end{equation}
where $g(x^i(t))=r_i(t) x^i(t)$ in the multiplicative case and 
$g(x^i(t))=r_i(t)$
in the additive case. We have used a small amplitude $\zeta=10^{-3}$ for the
noise, and $r_i(t)$ is a random number between $-1$ and $1$. It is chosen
anew for each map at each time step.

Fig. 9(a) and (b) show in a logarithmic scale the histograms of
distributions over pair distances in the glassy phase of the globally
coupled logistic map under two different choices of the initial conditions
for the same coupling intensity. We see that the resulting distributions are
very different. Note that both distributions have been averaged over
time, so that the peaks and irregularities in these figures reveal the
persistent structure of the underlying attractors. Fig. 9(c) shows how these
distributions are influenced by introducing into the GCM a weak additive
(dashed line) or multiplicative (bold line) noise of intensity $\zeta
=10^{-3}$ according to Eq. (7). The noises wash out the fine jagged
structure of the distribution and, more importantly, make it independent
of the initial conditions. The resulting distributions become then clearly
similar to the respective distribution we obtain at the same coupling
intensity for RCM, Fig. 9(d).

\begin{figure}
\centerline{\psfig{file=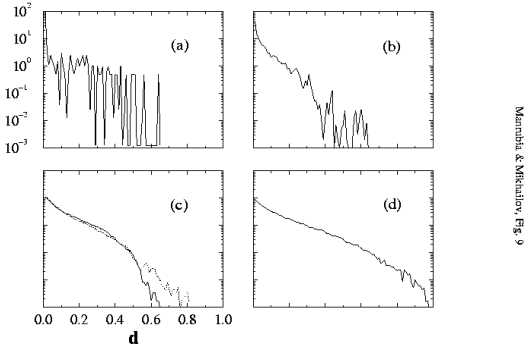,width=8.cm,angle=0}}
\caption{Normalized histograms over pair distances for a GCM of size 
$N=50$ and two different initial conditions (a,b), for the same GCM in the
presence of an additive (dashed line) or multiplicative (bold line) noise of
intensity $\zeta =10^{-3}$ (c), and for a randomly chosen
network of the same size with connectivity $\nu =0.8$ (d). The coupling
intensity is $\epsilon =0.45$ in all these plots; time averaging is
additionally performed.}
\end{figure}

We would like to emphasize that this parallelism between quenched
disorder and dynamical noise might hold in the intermittent phase, but
of course the mechanisms leading to the dynamical behaviour observed in
other phases cannot be (at least solely) ascribed to noise. In the 
clustering phase, for instance, it should be clear that
the fixed (though disordered) structure of the network plays the main role. 
In this sense, we have analysed the dynamical behaviour of the clusters 
by measuring in a fixed network how many elements belong to a certain 
cluster and how many clusters are formed at each time step. 
A cluster ${\cal C}_m(\delta ; t)$
for a given precision $\delta$ and at time $t$ is formed by $k_m$ 
elements, $m=1, \dots, M$ such that all of them have at least another
element of the cluster at a distance $d<\delta$, that is
$d=|x_i(t)-x_j(t)|<\delta$, $\forall i$ and some $j \in {\cal C}_m$ in
order to say that also $i \in {\cal C}_m$.

In Fig. 10 we represent the size of all clusters in a network with $N=50$ 
elements as a function of time (for a fixed precision
$\delta=0.1$). Four different values of $\epsilon$ in the clustering
phase have been chosen. In the first plot (Fig. 10(a), $\epsilon=0.25$), 
the elements tend to cluster but the groups are still relatively
unstable. A closer inspection of the clustering dynamics reveals that a
group of 19 elements keeps stable in time, while another cluster
containing
the 31 remaining elements splits often in subgroups of sizes (17,14),
(18,13) or (20,11) among others. If the coupling strength is increased,
also the stability of the clusters increases, and their average lifetime
becomes longer. In Fig. 10(b), for $\epsilon=0.28$ we observe that, in
fact, larger and more stable clusters are formed. Now the elements are 
divided into a cluster with 20 elements and a second group with 30 that
often splits into two subclusters with 27 and 3 elements, respectively.
Some irregularities in the dynamics are also found. In the hard-locking
phase for this network (Fig. 10(c), $\epsilon=0.30$) two stable clusters 
with 24 and 26 elements are formed. As already discussed, the maps display
periodic trajectories in this narrow parameter region. For a slightly
larger $\epsilon=0.32$, in Fig. 10(d), we see a first stable group with 
27 elements and a second one with 23 including a weakly coupled map 
(which periodically leaves the cluster). 

\begin{figure}
\centerline{\psfig{file=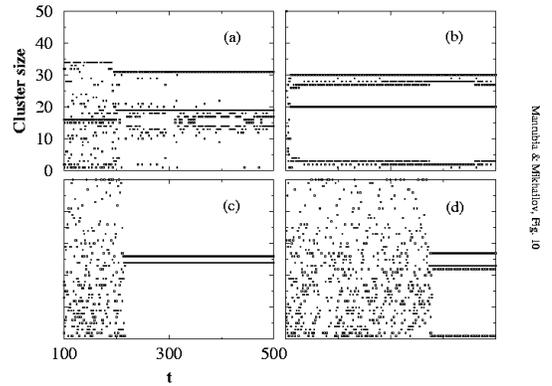,width=8.cm,angle=0}}
\caption{Size and stability of synchronous groups in the clustering phase.
The dynamical behaviour of a fixed network with $N=50$ elements and
$\nu=0.8$ is analysed for coupling strengths $\epsilon=0.25$ (a),
$\epsilon=0.28$ (b), $\epsilon=0.3$ (c), and $\epsilon=0.32$ (d). The
precision is in this case $\delta=0.1$. Weakly coupled elements coexist
with long-lived clusters which often split into smaller subgroups.}
\end{figure}

We have observed that the transient time required for the system to fall
into the clustering phase increases with the increase of $\epsilon$. At
the same time, it is seen that elements have a trend to condense in a
single group, as revealed by the presence of some time steps where
the cluster size equals $N$. This is not found to happen at the beginning 
of the clustering phase, for values of $\epsilon < 0.27$. 
Finally, for $\epsilon \approx 0.35$ (in the network of Fig. 10) the 
maps do not form clusters any more and the intermittent phase begins.

\section{Partitions into dynamical clusters}

In this section we more closely examine the structure of the dynamical
clustering phase in RCM. In this phase, groups of maps moving together in a
robust way and forming long-lived clusters have been observed.
The emerging cluster structure is biased by the connection patterns of the
underlying network. To demonstrate this, we introduce {\it relative
connectivities} that are defined below.

Let us suppose that at time $t$ and with precision $\delta $ our system
separates into $M$ clusters, ${\cal C}_{m}(\delta;t)$, $m=1,\dots ,M$, each
of which containing $k_{m}$ elements. The relative connectivity inside a
cluster $m$ is then defined as

\begin{equation}
\nu _{eff}^{m}=\frac{1}{\nu k_{m}(k_{m}-1)}\sum_{\langle ij\rangle
}T_{ij},\;\;\;i,j\in {\cal C}_{m},\;\;\;\;i\ne j
\end{equation}
where the sum is taken over all pairs of elements belonging to this cluster.
The relative connectivity between two different clusters $l$ and $n$ is
given by
\begin{equation}
\nu _{eff}^{l,n}=\frac{1}{\nu k_{l}k_{n}}\sum_{\langle ij\rangle
}T_{ij},\;\;\;i\in {\cal C}_{l},\;\;j\in {\cal C}_{n}
\end{equation}
Thus defined, the relative connectivities are equal to $1$ if the
characteristic connectivity inside a cluster or between two clusters are
exactly the same as the average connectivity $\nu $ of the entire network.
Positive deviations of the connectivity inside a cluster ($\nu _{eff}^{m}>1$)
indicate that this cluster contains elements which are more strongly
connected than on the average. Respectively, when the relative connectivity
between two clusters is decreased ($\nu _{eff}^{l,n}<1$), this shows that
the elements belonging to these two separate clusters are less connected
than on the average and {\it vice versa}.

We can also define the average {\it intercluster} relative connectivity of
the entire network

\begin{equation}
\nu _{eff}^{Inter}={\frac{1}{M(M-1)}}\sum_{n,l=1;n\ne l}^{M}\nu _{eff}^{l,n}
\label{nueff}
\end{equation}
and its average {\it intracluster} relative connectivity

\begin{equation}
\nu _{eff}^{Intra}={\frac{1}{M}}\sum_{m=1}^{M}\nu _{eff}^{m}.  \label{nueff2}
\end{equation}

To characterize the cluster structure of the partially ordered phase, we fix
the coupling intensity $\epsilon $ and consider the state of the whole
system at a given time moment. By varying the precision $\delta,$ we obtain 
a hierarchy of cluster partitions, as seen with different resolutions. For
each resolution level, its relative connectivities are then calculated.
Fig. 11 presents the emerging hierarchical structure of dynamical cluster
partitions for a system of $N=50$ elements with coupling strength 
$\epsilon =0.23$ at three different precisions $\delta$. The numbers between
brackets correspond to the number of elements in each particular cluster.
The numbers inside the clusters are their relative connectivities and the
numbers on the links between the clusters yield the relative connectivity
between them. Though this pattern refers to a particular time moment (after
a long transient), it remains fairly stable in time. We
see that, as the precision $\delta $ is improved, the clusters split into
smaller subclusters, revealing a hierarchical tree-like structure\cite{Foot}. 
It can
also be observed in Fig. 11 that the relative connectivities inside a cluster
exceed one, whereas the relative connectivities between the clusters are
typically smaller than one. This indicates that the partition into dynamical
clusters is biased by the pattern of connections in the underlying network,
i.e. the elements belonging to a same cluster would generally have more
connections inside this cluster than with the elements belonging to other
dynamical clusters.

\begin{figure}
\centerline{\psfig{file=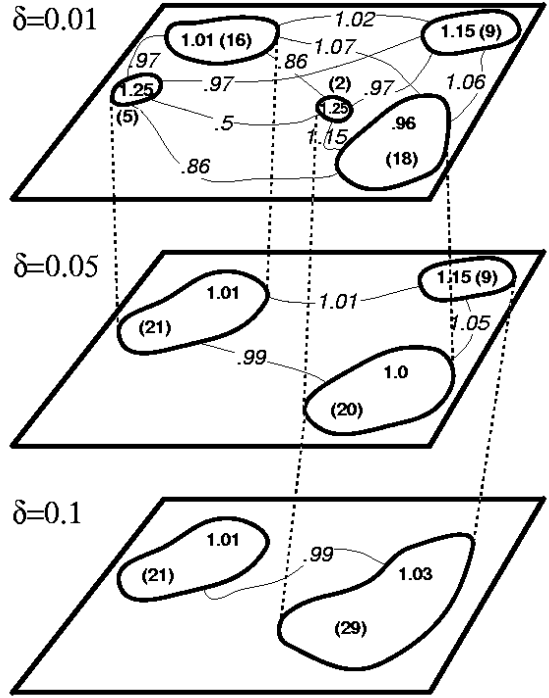,width=8.cm,angle=0}}
\caption{Hierarchical structure of dynamical clusters for a randomly chosen
network of size $N=50$ with the connectivity $\nu =0.8$ at $\epsilon =0.23$.}
\end{figure}

To check more accurately this suggestion, we have calculated inter- and
intracluster connectivities $\nu _{eff}^{Intra}$ and $\nu _{eff}^{Inter}$
for a larger system with $N=250$ and $\nu =0.8$ at $\epsilon =0.3$ with the
precision $\delta =0.1.$ These properties were averaged over $10^{4}$
independent graphs. The average intracluster connectivity was in that case $%
\left\langle \nu _{eff}^{Intra}\right\rangle =1.013(1),$ that is, slightly
higher than numerically generated average connectivity, $\left\langle \nu
_{eff}\right\rangle =1.0000(1)$. The average intercluster connectivity was $%
\left\langle \nu _{eff}^{Inter}\right\rangle =0.987(1)$ and thus lying below
$\left\langle \nu _{eff}\right\rangle $. Figure 12 shows the normalized
probability distribution over intracluster (solid line) and
intercluster (dashed line) connectivities in the studied ensemble of $10^{4}$
independently generated graphs. The maxima of the two distributions are
slightly shifted. But, perhaps even more important, we see that the
distribution of the intercluster connectivities is significantly broader
and has a wide shoulder extending towards lower connectivities.

\begin{figure}
\centerline{\psfig{file=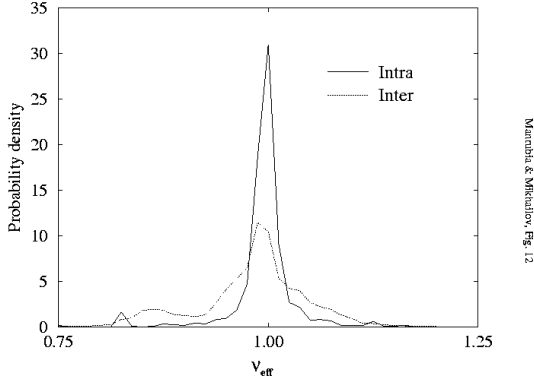,width=8.cm,angle=0}}
\caption{Statistical distributions over intracluster (solid line) and
intercluster (dashed line) effective connectivities in the clustering
partitions in an ensemble of $10^{4}$ independently generated random
networks of size $N=250$ with mean connectivity $\nu =0.8$ and coupling 
strength $\epsilon =0.3$. In this case, we have taken the precision 
$\delta =0.1$.}
\end{figure}

\section{Discussion}

Our numerical analysis reveals that mutual synchronization and dynamical
clustering represent a typical and robust form of collective dynamics in
random networks of coupled chaotic elements. The mutual synchronization
remains possible when almost half of all potential connections between the
elements are deleted and the dynamical clustering may be found even for the
more sparsely connected networks. For very low connectivities, we have seen
that the system could no longer synchronize and the dynamical clusters in
the partially condensed phase became less stable, i.e. their lifetimes were
getting shorter.

The different dynamical phases of RCM have been specified and compared with
their counterparts in GCM. The essential differences have been noticed in
the cluster structure and in the dynamics, as well as in the dependence on the
initial conditions. A rich clustering structure, depending on the network
architecture, was observed in RCM. The analog of the glassy phase of GCM was
however not found in the investigated randomly coupled maps, i.e. we have
not seen that the final attractor depended on the initial conditions for any
set of parameters. In this sense, the quenched disorder of the random
network appears to play a role similar to that of noise in this phase.
It might also be that the transition from synchronization to the intermittent
phase in RCM would admit a characterization in terms of a blowout
bifurcation \cite{SO}, and that the intermittent phase that we observe
immediately after the synchronous state be in fact a case of on-off
intermittency. This picture would be 
consistent with our numerical results and with the fact that the disorder 
in the network architecture destroys the degeneracy of the dynamical matrix 
in GCM \cite{Kan3} and generates a whole hierarchy of Lyapunov exponents.

Though our investigations were made only for networks formed by coupled
logistic maps, similar results would probably hold for networks made of
other chaotic maps or elements with continuous chaotic dynamics. Indeed, the
behavior in globally coupled logistic maps strongly resembles what is found
in various globally coupled populations of chaotic dynamical systems \cite
{D1,M1}.

The systematic study of RCM implies the analysis of the behavior of the
system under the change of four relevant parameters: The average connectivity 
$\nu$, the coupling strength $\epsilon$, the parameter of the individual
map $a$, and the system size $N$. Our main interest in this study was to
introduce Randomly Coupled Maps and to give some insights on the role of
the network architecture in the dynamics. Hence, we have mainly
investigated the two parameters $\nu$ and $N$, and reanalysed the known
phases for GCM when $\epsilon$ varies from zero to unity. Many of our 
investigations were performed with control parameter $a=2$ of the
logistic map. This value is somewhat special, since at $a>2$ the
trajectories become infinite and the chaotic attractor disappears in a
boundary crisis \cite{All}. Other simulations for smaller values of the 
control parameter $a$ show a similar qualitative behaviour.

In our study, the networks were generated by independently choosing with a
certain fixed probability the connection between any two elements. Thus
constructed, the connection patterns are random, but statistically uniform.
We have analysed systems with sizes ranging from a few elements to 
$N=2^{11}$. Generally, the synchronization threshold depended not only on the
system size and on the average connectivity, but also on the particular
architecture of a chosen network. We have seen, however, that variations in
the synchronization threshold for networks with the same mean
connectivity became much weaker when the network size was increased, and
that the distance to the mean field threshold given by GCM had the
functional form $(\epsilon^*-\epsilon^*_{GCM}) \simeq N^{-1/2}$.
The existence of a universal synchronization threshold in such 
randomly generated networks in the limit $\nu N \to \infty$ is thus 
expected. The statistical uniformity, introduced in this paper through 
the independent choice of individual connections, is a special feature that 
should not necessarily be present in complex networks. Natural networks may 
have various topological structures \cite{Watts} which can also result from 
evolutionary processes \cite{Kauff,Kim}. It would be interesting to
see how synchronization and the dynamical clustering phenomena are
influenced by such structures.

We have found that the network architecture biases the partitions of the
network into dynamical clusters and determines interactions between the
clusters which lead to their collective dynamics. This puts forward the task
of {\it engineering} the networks with the desired dynamical clustering
properties. One can apparently design systems that would display an
arbitrarily chosen partition into several exactly synchronized clusters (see
the Appendix). The collective dynamical behavior can represent an important
practical function of a network. The evolution of a network, proceeding
through random mutations, may then be guided towards the optimization of its
collective dynamics. Indeed, examples of dynamical networks that evolve to
reproduce given temporal `melodies' have already been constructed\cite
{melody,melody1}. We want therefore to emphasize that the evolution
of networks can also be steered to reach better synchronization properties
or to approach a certain dynamical clustering structure.

Finally, we note that when the dynamical clustering is taking place,
coherent clusters can be interpreted as super elements that form an emerging
dynamical network of a higher structural level. Taking into account the
large variety of clustering partitions and their sensitivity to the coupling
intensity, RCM systems may thus be viewed as a living space that supports
different dynamical (meta)networks and may retrieve a particular such
network under appropriate changes of the control parameters.

\section*{Acknowledgments}

The authors acknowledge interesting discussions with Dami\'{a}n H. Zanette.
SCM gratefully acknowledges the support from the Alexander von Humboldt 
Foundation (Germany).

\section*{Appendix}

It was earlier noted that the network architecture influences the critical
coupling intensity $\epsilon^{*}$ at which synchronization first appears 
and favours certain preferred partitions
of elements into dynamical clusters. In this Appendix we analyze the role of
the network architecture in the dynamical clustering phenomena for small
networks consisting of only $N=5$ elements. If the network connectivity is
fixed at $\nu =0.6$, there are just four such networks shown in Fig. 13. We
have systematically investigated their synchronization properties for
various values of the control parameter $a$ of the logistic map in the
interval from $1.42$ to $2$ with increment $\Delta a=0.02$ and for the
coupling intensity $\epsilon$ in the interval from $0$ to $\nu$ with 
increment $\Delta \epsilon =0.01$. In this case we have considered that
two maps are synchronized if they have exactly the same state. This is
now licit because of the high degree of symmetry of the networks. 
The percent of parameter pairs leading to
each of the possible clustering configurations is displayed in Table 1.
In Fig. 13, elements with the same symbol synchronize (i.e. they form a 
cluster) with the higher probability. Different symbols stand for
different clusters.
The value of $\epsilon$ at which the elements in each of the networks
synchronize can take a wide spectrum of values. For instance, for $a=1.6$
it changes from 0.56 (case(A)) to 0.96 (case(B)) \cite{comment}. Moreover, 
we have found that $\epsilon^*$ is a non-monotonic function of $a$, and can
be even decreasing depending on the graph. Hence, if $N$ is small, each 
network has to be treated independently (as in the example here analyzed).

We see that indeed in the majority of cases the clustering partition follows
the pattern of connections in the graph. Moreover, some of the potential
highly asymmetrical partitions have never been observed (like the partition
into (1,5) (2,3) (4) for the graph A). This shows that the symmetry of
connections inside a graph plays an important role in the dynamical
clustering phenomena.

Looking at the graph A, we see that the dynamical equations of its elements
are not changed under the relabeling $(1,2,3,4,5)\to (2,1,4,3,5)$, reflecting
the symmetry with respect to permutations in the matrix $T_{ij}$ for this
graph. It seems highly plausible that synchronous clusters would generally
be much easily formed by {\it indistinguishable} elements, defined as those 
whose dynamical
equations are identical under permutations. This is the reason that
allows synchronization to be of the hard locking (exact) type in this case. 
The numerical results shown in Table I support this statement.

Consider, for example, the situation in which the clusters (1,2), (3,4) and
(5) have been formed in (A). Let us call $x^1(t)=x^2(t) \equiv x$, 
$x^3(t)=x^4(t)
\equiv y$ and $x^5(t) \equiv z$ and write down the dynamic equations for the
new 3-cluster system:

\[
x(t+1)=\left( 1-{\frac{5}{12}}\epsilon \right) f(x)+{\frac{5}{12}}\epsilon
f(y)
\]
\[
y(t+1)=\left( 1-{\frac{5}{6}}\epsilon \right) f(y)+{\frac{5}{12}}\epsilon
\left( f(x)+f(z)\right)
\]
$$
z(t+1)=\left( 1-{\frac{5}{6}}\epsilon \right) f(z)+{\frac{5}{6}}\epsilon f(y)%
\eqno(A.1)
$$
On the one hand, we do not observe any further (at least trivial) symmetry
in these equations. On the other hand, the problem of global synchronization
in our original graph has moved to the problem of synchronization of
asymmetrically connected oscillators which moreover have different values
for the coupling strength. According to this result, we believe that the
problem of synchronization and clustering in RCM might also include
asymmetrically connected networks and probably some cases of coupled systems
with a distribution $\Pi (\epsilon )$ of $\epsilon $ values \cite
{Dam,Stiller}.

\begin{figure}
\centerline{\psfig{file=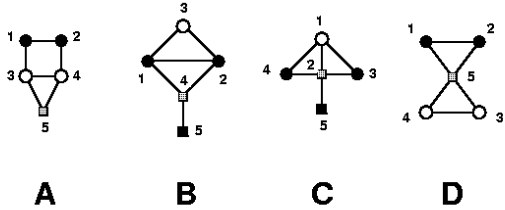,width=8.cm,angle=0}}
\caption{Four different possible configurations of a network with $N=5$
elements and $\nu =0.6$. In each graph, favoured synchronizations are
displayed using the same symbol.}
\end{figure}

\onecolumn

\begin{table}[tbp]
{TABLE I. Percent of realizations (\%) for the clusterings shown and for
each of the networks in Fig. 13. }
\begin{tabular}{ccc|ccc}
Network & Clusters & \% \hspace{1cm} & Network & Clusters & \% \\ \hline
{\bf A} & (1,2,3,4,5) & 5 \hspace{1cm} & {\bf B} & (1,2,3,4,5) & 1.5 \\
& (1,2,5) (3,4) & 2 \hspace{1cm} &  & (1,2,4) (3) (5) & 1 \\
& (1,2) (3,4,5) & 1 \hspace{1cm} &  & (1,2) (3) (4) (5) & 75 \\
& (1,2) (3,4) (5) & 59 \hspace{1cm} &  & (1,2,3,4) (5) & 1 \\
& turbulent & 33 \hspace{1cm} &  & turbulent & 21.5 \\ \hline
{\bf C} & (1,2,3,4,5) & 2 \hspace{1cm} & {\bf D} & (1,2,3,4,5) & 4 \\
& (1,2,3,4) (5) & 1 \hspace{1cm} &  & (1,2,3,4) (5) & 1 \\
& (1,3,4) (2) (5) & 17 \hspace{1cm} &  & (1,2) (3,4) (5) & $57$ \\
& (1,3,4) (2,5) & 0.5 \hspace{1cm} &  & (1) (2) (3,4) (5) & 1.5 \\
& (1,2) (3,4) (5) & 0.5 \hspace{1cm} &  & (1,2) (3) (4) (5) & 1.5 \\
& (1) (2) (3,4) (5) & 17 \hspace{1cm} &  & turbulent & 35 \\
& turbulent & 62 \hspace{1cm} &  &  &
\end{tabular}
\end{table}


\begin{references}
\bibitem{Foot} In Fig. 10, where the temporal behaviour of clusters is 
shown, only two or three clusters can often be seen. This however is explained 
by the low precision ($\delta=0.1$) used in these plots. When the precision is
increased, splitting of the clusters would be observed, similar to what
is shown in Fig. 11.
\bibitem{comment} The value of $\epsilon$ is theoretically bounded by
$\nu$: For $\epsilon > \nu$ divergencies in the values of $x^i$ might be
found. In practical terms, however, it is found that interactions among
the elements prevent their states from attaining extremal values, and
thus values of $\epsilon > \nu$ can be used with caution.
\end{references}
\end{document}